# Adaptive recycled plastic architecture: Vacuum-Sealed Chainmail Structures Through Computational Design

Yi Xu, Farzin Lotfi-Jam and Mustafa Faruki

## Abstract

The construction industry is a major consumer of raw materials, accounting for nearly half of global material usage annually, while generating significant waste that poses sustainability challenges. This paper explores the untapped potential of recycled plastics as a primary construction material, leveraging their lightweight, flexible, and customizable properties for advanced applications in modular chainmail systems. Through a computational workflow, the study optimizes the design, testing, and fabrication of vacuum-sealed chainmail structures composed of recycled plastic filaments, demonstrating their adaptability and structural performance for architectural use.

Key contributions include a novel methodology for integrating recycled plastic filaments into chainmail geometries, validated through 2D sectional testing, 3D shell structure generation, and physical modeling under vacuum constraints. The research identifies the rectangular chainmail configuration as the most efficient and adaptable, achieving superior deformation capacity, material efficiency, and load-bearing performance. Optimization strategies for temporary structures highlight practical deployment potential, balancing material savings, usable area, and water drainage efficiency.

The findings offer a foundation for innovative applications in extreme conditions, including disaster-prone areas, high-altitude environments, underwater platforms, and extraterrestrial habitats. These applications leverage the lightweight, adaptable, and durable properties of recycled plastics and modular chainmail systems, bridging the gap between waste management and high-performance design while addressing unique challenges in harsh and resource-constrained environments.

# Introduction

## Motivation

For centuries, architecture has been inspired by nature, utilizing biomimicry to imitate biological efficiency in design and material innovation (1). Prominent examples encompass Antoni Gaudí's catenary structures (2), Frei Otto's lightweight tensile systems (3), and Buckminster Fuller's geodesic domes (4)—all of which investigate structural optimization based on natural principles. Those methodological approaches are not limited to form imitation but extend to material selection and application, extracting high-performance characteristics from nature or experimental materials and adapting them to architectural needs.

Simultaneously, progress in materials science has propelled the utilization of performance-oriented material systems in architecture, transcending conventional construction constraints. This study adheres to the methodological principles of biomimicry; however, it concentrates on high-performance materials created in laboratory environments rather than examining natural forms. It specifically examines the architectural adaptability of vacuum-sealed chainmail structures, a system demonstrated to possess variable stiffness, flexibility, and outstanding mechanical properties (5). The primary aim is to examine how this experimental material can be modified and incorporated into architectural structures, fulfilling requirements for structural integrity, sustainability, and practical application.

This methodology corresponds with conventional biomimetic objectives of deriving efficient structural principles and optimizing them for constructed environments. Instead of utilizing natural forms, it employs experimental materials as the basis for innovation. Material sustainability is increasingly central to biomimetic architectural strategies, especially as global issues like plastic waste necessitate innovative solutions.

At the same time, the global plastic waste crisis presents an urgent challenge and an opportunity for architectural innovation. In 2020, the U.S. Environmental Protection Agency (EPA) reported that plastic waste generation had reached 35.7 million tons, yet only 4.5% was successfully recycled, underscoring severe inefficiencies in waste management (6). Unlike industries such as food packaging and medical applications, where material purity is strictly regulated (7), the construction sector prioritizes durability, cost-effectiveness, and adaptability, making it particularly suited for integrating recycled materials (8).

While prior studies have confirmed the mechanical feasibility of vacuum-sealed chainmail fabric, systematic research on its material sustainability, architectural applicability (5), and real-world implementation remains lacking. Chainmail structures, characterized by unique geometric and mechanical properties, have yet to be extensively explored within the architectural domain. To address this gap, this study proposes the integration of vacuum-sealed chainmail fabrics with recycled plastics, specifically investigating their structural adaptability and sustainability potential in architectural applications.

By integrating computational workflows, structural simulations, and material optimization, this study not only explores the mechanical feasibility of chainmail-based structures but also establishes a scalable methodological framework for adapting experimental materials into sustainable architectural solutions. This strategy enhances structural performance optimization and material sustainability, showcasing the viability of recycled plastics in high-performance architectural systems and providing a novel technological path for circular economy applications.

## Related Work

*State of the Art*

### Chainmail Systems: Historical Legacy and Modern Potential

Historically, chainmail has been regarded for its flexibility and durability as a form of defensive armor (9). The construction—interlinked rings of iron, steel, or bronze—facilitated enhanced mobility relative to rigid plate armor while offering adequate protection against slashing weapons (10). Chainmail's capacity for dispersing forces among its interconnected rings diminished the probability of penetration.

Contemporary advancements have transformed chainmail systems through the integration of distinctive materials and technologies, hence augmenting their functionality beyond traditional armor applications. Recent research, especially in soft robotics, illustrates their effectiveness in applications necessitating programmable flexibility. Actuated chainmail fabrics can shift between stiff and flexible states with embedded shape memory alloy (SMA) actuators, facilitating dynamic reactions to external stimuli (11). This discovery facilitates the creation of reconfigurable orthoses, rigid clothing, and load-bearing structures, greatly expanding their range of applications.

Furthermore, chainmail systems in vacuum-sealed topologies attain stiffness enhancements of up to 25 times, demonstrating remarkable mechanical qualities under pressures reaching 2 MPa (11). Their lightweight, hollow 3D-structured particles, possessing densities as low as 0.18 g/cm³, improve mobility and versatility, rendering them appropriate for wearable robotics and dynamic protection equipment. Moreover, their capacity to conform to doubly curved surfaces and compact folds render them formidable contenders for adaptive architectural systems (12).

These attributes offer chainmail an appropriate choice for adaptive architectural systems, wearable robotics, and dynamic safeguards. Nonetheless, contemporary research is still nascent, concentrating on enhancing material efficiency and investigating scalable production methods.

## RPET as Filaments for 3D Printing

The combination of 3D printing technology with waste repurposing presents a revolutionary option for the circular economy, integrating sustainable development objectives with new material uses (13). In 2024, Hassan et al. indicates that recycled polymers utilized as 3D printing materials demonstrate durability, potentially beyond that of virgin materials (14). Incorporating 30% by weight of glass fibers into polyethylene terephthalate (PET) improves its strength and heat deflection temperature (15). In recent years, rPET filaments have drawn interest for their multiple applications in sustainable building, showcasing considerable potential as environmentally acceptable substitutes for conventional materials (16). PET constitutes 10%-12% of global plastic production and is widely used for its durability and excellent recyclability (17).

Due to developments in recycling technology, the construction sector has progressively embraced rPET, especially in the production of intricate geometric shapes and functional components. The Nanjing Happy Valley East Gate project employed high-precision 3D printing to transform rPET and modified plastics into facades, decreasing construction time and expenses while minimizing material waste by 30%-50%. The printed components acquired a recycle rate of up to 90% (18). Similarly, Wong in 2018 presents the 'Cloud Village' at the Venice Biennale demonstrated the versatility of rPET in creative constructions, utilizing 3D printing techniques to produce modules that reduced material waste and facilitated disassembly and reuse (19). In addition, the DB Schenker Upcycling Hub in Singapore exploited additive manufacturing and digital design technologies to convert 30,000 PET waste products into chandeliers and furniture components (20). These instances illustrate the significant potential of rPET, offering robust support for a more environmentally friendly future in the building sector.

In summary, rPET filaments exhibit the capacity of recycled materials to transform construction methodologies. By improving mechanical properties, minimizing environmental impact, and facilitating creative uses, it adheres to the principles of a circular economy and establishes a robust foundation for resource-efficient and environmentally responsible construction processes.

*Research Gap and Proposed Approach*

### Recyclability Challenges in Existing rPET Applications

Recent research on rPET predominantly examines its incorporation into concrete and cementitious composites, either as an additive to augment tensile and flexural strength (21,22) or to enhance durability within cement matrices (23). Nevertheless, these applications frequently entail chemical bonding, which undermines recyclability by complicating material separation at the end of their lifecycle. As a result, these approaches do not facilitate a circular material system.

In contrast, this study introduces a mechanically assembled rPET-based chainmail system, wherein individual filaments preserve their physical integrity, enabling direct remelting and reuse without degradation. This method adheres to circular economy principles by removing chemical bonding, while maintaining the mechanical and structural benefits of chainmail systems.

### Performance of Complex Geometries in Chainmail Systems

Current research on chainmail systems predominantly focusses on single axis bending and uniform compression, neglecting arbitrary rotations and complex curvatures. Investigations into bending properties (24) and stiffness tunability (5) have evaluated reactions to external forces but are limited to predetermined deformations. Tang et al. created computational models for interlocking materials; however, they did not consider freeform curvature transformations essential for architectural applications (12).

Vacuum-sealed chainmail systems are inadequately investigated regarding stiffness variations and deformation mechanics under dynamic loads. Enhancing multi-directional adaptability and material-driven deformation control will improve architectural viability, establishing one direction for sustainable, adaptive applications.

### Practical Applications in Post-Disaster Relief

Post-disaster rapid-deployment shelters necessitate lightweight, adaptable, and reusable materials; nevertheless, achieving a balance between flexibility and structural integrity poses a significant challenge. Although inflatable membranes, tensegrity frameworks, and modular panels (25,26) have been investigated, they all exhibit constraints.

Membrane and tensile-based shelters are portable and compact; however, they frequently lack the rigidity necessary to endure severe conditions such as high winds, seismic events, and substantial snowfall (26). Conversely, prefabricated modular panels provide stability; instead, their transportation and assembly limitations diminish their practicality in rapid-response operations (5).

The suggested chainmail-based system presents a feasible choice by achieving a balance between adaptability and structural integrity. The vacuum-sealed structure increases rigidity as required, while the modular design facilitates efficient transportation and assembly. This methodology seeks to enhance deplorability and resilience in post-disaster shelter solutions.

This study employs a systematic methodology to investigate rPET-based chainmail systems in vacuum-sealed environments. The methodology seeks to evaluate structural efficacy, adaptability to intricate geometries, and prospective applications in post-disaster contexts. A multi-phase computational workflow is employed to analyses key parameters including solid-to-gap ratios, deformation behaviors, and geometric adaptability. This section delineates the procedures undertaken to design, evaluate, and enhance these systems for architectural purposes.

# Methodology

## General Overview of Computational Design Workflow

This research employs a multi-phase methodology (Figure 1) to design, evaluate, and enhance chainmail systems for architectural uses within vacuum-sealed parameters. The procedure commences with the Chainmail System Setup, establishing essential parameters including solid-to-gap ratios and uniform part dimensions to guarantee structural equilibrium and comparability. 2D Sectional Testing investigates deformation behaviors and curvature adaptability, establishing a foundation for advancing to 3D Shell Structure Generation, where amplitude and frequency parameters influence three-dimensional geometries for structural evaluation.

The following stages concentrate on enhancing and confirming the designs. The Filtering Process removes redundant geometries based on perimeter and surface area measurements, while Physical Modelling and Vacuum Constraints validate computational outcomes through scaled prototypes in vacuum-sealed environments. Ultimately, Load Analysis for Free-Form Shell Structures and Form and Material Optimization guarantee that the designs satisfy structural integrity, material efficiency, and functional performance criteria, thereby confirming their viability for practical applications.

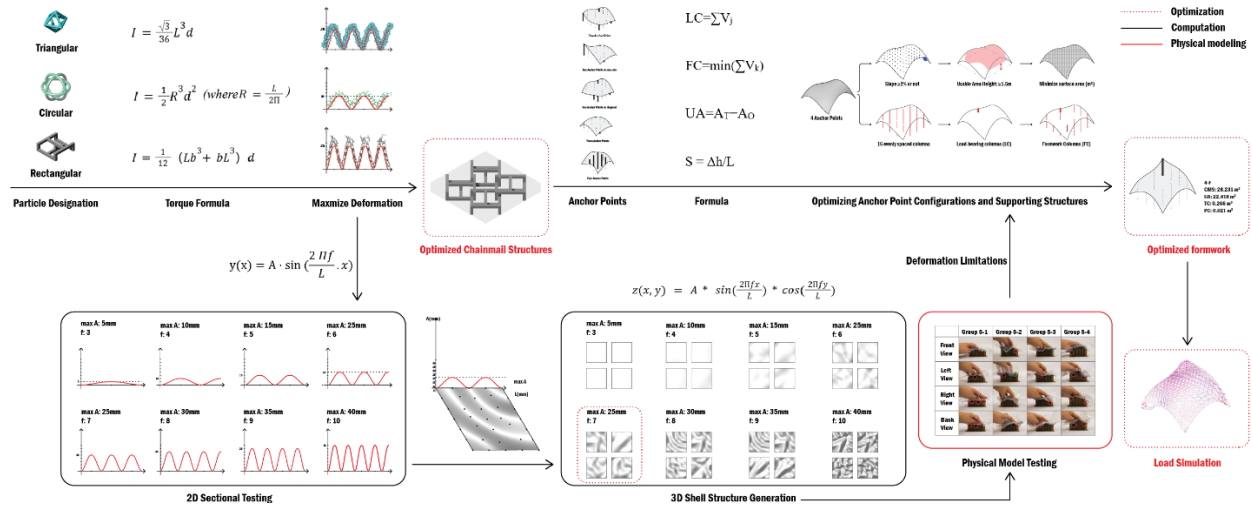

Figure 1: General overview of the computational design workflow.

## Chainmail System Setup

The design of triangular, circular, and rectangular chainmail systems (Figure 2) incorporates amplitude and frequency characteristics into a coherent and systematic workflow. Two fundamental constraints—a solid-to-gap ratio of 8% and uniform part diameter—were implemented to achieve structural balance, comparability, and scalability.

|          | Triangular | Circular | Rectangular |
|----------|------------|----------|-------------|
| L/b (mm) | 7.5        | 11       | 11          |
| d (mm)   | 1          | 1        | 1           |

Figure 2: Geometric Configurations and Dimensions of Chainmail Units.

- The 8% solid-to-gap ratio (R*sg*) enhances flexibility, load-bearing capacity, and topological interlocking, while reducing localized failures. Modular modules, created in Grasshopper, sustained R*sg* = 0.08 by recurrent simulations to ensure uniform force distribution. R*sg* is defined as:

$$Rsg = \frac{Vs}{Vg + Vs}$$

  Where:

  $V_s$: Total solid area in a unit cell,

  $V_g$: Total gap area in a unit cell.

- A uniform diameter d = 1mm removes size-dependent variables, hence isolating deformation behaviors particular to geometry. This standardization guarantees equitable performance evaluations, assigning variations only to geometric forms (e.g., triangles, circles, rectangles) and their distribution patterns.

| Triangular units | $I = \frac{\sqrt{3}}{36} L^3 d$ | L = 7.5mm, d = 1 mm |
|---|---|---|
| Circular units | $I = \frac{1}{2} R^3 d^2$ (where $R = \frac{L}{2\Pi}$) | L = 11 mm, d = 1 mm |
| Rectangular units | $I = \frac{1}{12}(Lb^3 + bL^3) d$ | L = b = 11 mm, d = 1 mm |

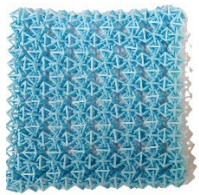
Triangular

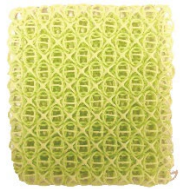
Circular

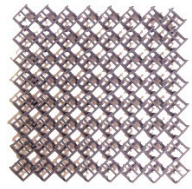
Rectangular

Figure 3: 3D-Printed Samples of Triangular, Circular, and Rectangular Chainmail Structures.

These constraints provided a streamlined framework for exploring deformation behaviors while optimizing structural and material efficiency, forming the foundation for further testing and refinement.

## 2D Sectional Testing

Two-dimensional sectional testing was performed to assess deformation characteristics and curvature adaptability under regulated settings. This method diminished the intricacy of managing 3D shapes while facilitating a direct evaluation of structural effectiveness.

The testing framework utilized scaled models to represent real-world scenarios. Each 2D section (Figure 4) corresponded to a true length of 2m, scaled down to 1:20 for practical testing. Key parameters, including amplitude and frequency, were systematically varied:

- Amplitude (A): Wave heights began at 0 mm and increased in 5 mm increments up to 40 mm, allowing for the observation of maximum deformation capacity across triangular, circular, and rectangular geometries.

- Frequency (f): Frequency was defined by dividing the 2m length into equal segments based on grid counts. Starting from a grid count of 3 (3×3), increasing in intervals of 1, with proportional adjustments to amplitude to simulate real-world deformation scenarios.

The deformation profiles were described using the sinusoidal function:

$$y(x) = A \cdot \sin\left(\frac{2\Pi f}{L} \cdot x\right)$$

where:

- y(x): deformation height at any point x,
- L: total length of the tested section (2000 mm in real-world scale).

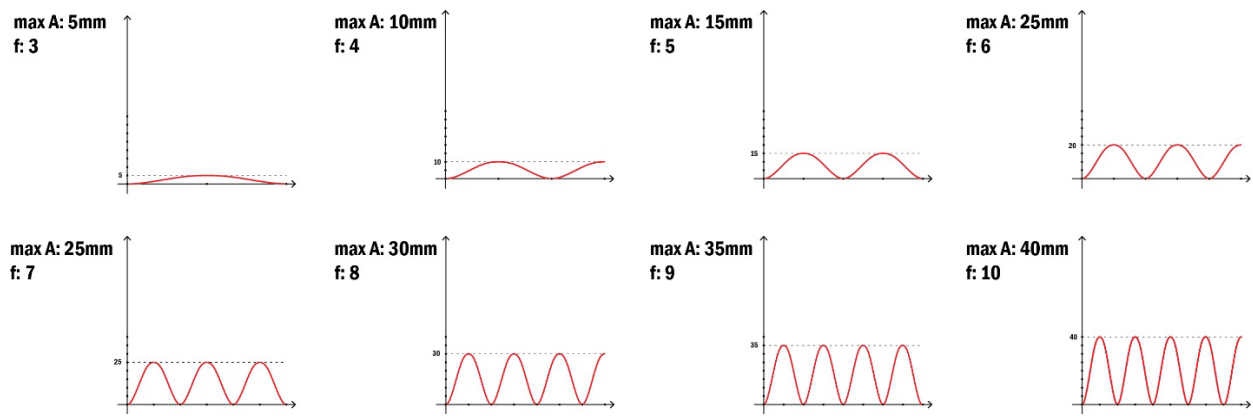

Figure 4: Amplitude and Frequency Correlations in Rectangular Chainmail Deformation Profiles.

## 3D Shell Structure Generation

Building on the results from 2D Sectional Testing, the most optimal chainmail structure—characterized by its maximum adaptable curvature—provides the A and f values as a foundation for generating 3D shell structures (Figure 5). The X- and Y-axis grids maintain the same logic, where the number of control points is determined by $F^2$ (e.g., 3*3 for F=3).

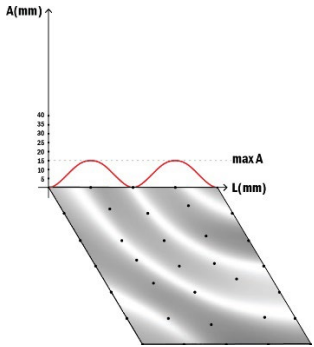

Figure 5: Amplitude and Frequency Distribution in Rectangular Chainmail Shell Deformation.

In this phase, the Z-axis becomes the key variable, with the range of Z-axis movement for each control point directly constrained by the corresponding A from the 2D tests. For example, if A=25 mm, the Z-axis range for each control point is limited to $0 \leq Z \leq \frac{A}{5}$. The deformation for any given control point can then be expressed as:

$$z(x, y) = A * \sin(\frac{2\Pi fx}{L}) * \cos(\frac{2\Pi fy}{L})$$

For each combination of A and F, 20 iterations are generated by introducing random variations in the Z-axis positions. This allows for a systematic exploration of deformation patterns in three-dimensional geometries while leveraging the optimal parameters established in the 2D analysis.

## Filtering Process

A filtering process was designed to minimize redundancy and retain significant variation in the extensive dataset produced from each combination of amplitude (A) and frequency (f). The aim was to preserve only four distinct forms for each combination.

Figure 6 depicts the filtering process that assesses iterations according to Perimeter (P) and Surface Area (A). Geometries are deemed distinct if they above the tolerance limits ΔP or Δa, which are specified as follows:

$$|P1-P2| > \Delta P \quad \text{or} \quad |a1-a2| > \Delta a$$

This ensures that only iterations with significant differences in perimeter or surface area are selected.

Utilizing Group 5 as an example, a tolerance was implemented to discover and exclude forms with negligible differences, isolating only four unique geometries. This method systematically reduced the dataset while preserving diversity, ensuring an efficient examination of the further study.

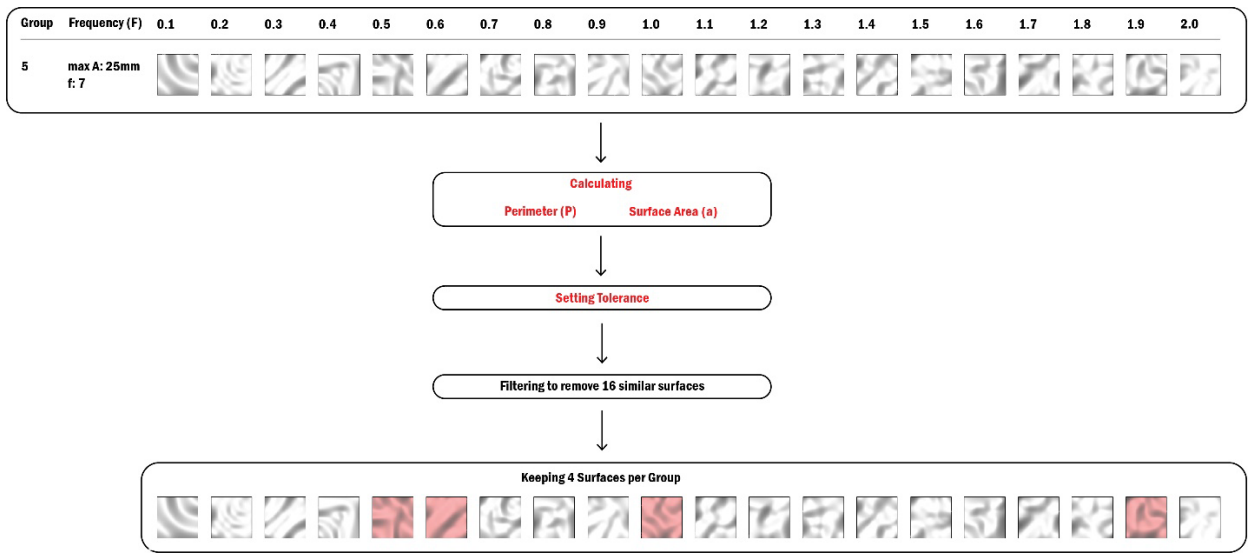

Figure 6: Surface Selection Workflow: Filtering Process for Optimizing Shell Structures.

## Physical Modeling and Vacuum Constraints

Given the transition from 2D testing to 3D forms and the addition of a plastic membrane, the maximum achievable deformation of the chainmail structures may differ from the computational predictions. To address this, 1:20 scaled physical models were created to provide a visual assessment of their behavior under vacuum-sealed conditions.

This phase concentrated on the four filtered iterations. Meanwhile, those models were enclosed in vacuum-sealed plastic bags to replicate practical conditions, facilitating an analysis of their deformation characteristics and structural adaptability. Four perspectives—front, back, left, and right views—were captured to verify the structural integrity. Take Group 5-1 (Figure 7) as example, as detachment of rods or gaps in sealing, were marked and analyzed.

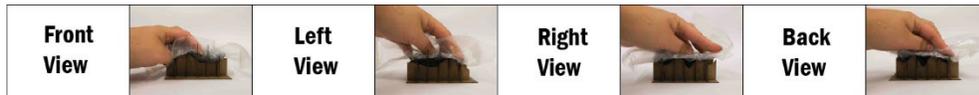

Figure 7: Vacuum-Sealed Rectangular Chainmail System: Multi-Angle Physical Model Testing Views.

Load Analysis for Free-Form Shell Structure

Building on the insights from physical modeling and vacuum constraints, the load-bearing capabilities of rectangular chainmail structures were further analyzed to evaluate their performance under practical conditions. The vacuum-sealed prototypes, fabricated at a 1:20 scale revealed deficiencies including deformation and compression. These findings emphasized the necessity of quantitatively understanding how such structures respond to external forces.

To complement the testing, a 2m*2m free-form shell structure made of recyclable plastic (rPET) was selected for computational load analysis (Figure 8). This structure is intended as an alternative to traditional reinforced concrete shell structures, and its analysis was conducted using standardized guidelines commonly applied to concrete structures. The evaluation focused on critical load types—dead, live, snow, and wind—and their combined effects, validating the structural integrity and feasibility of the proposed material under practical conditions.

The vacuum-sealing process could not be directly recreated in Karamba3D. To approach this scenario, the outermost points of the chainmail structure were excised to construct a mesh connecting the membrane to the inside chainmail structure. A compression force of 1N was exerted on the membrane to more accurately replicate vacuum sealing. This method, however, has constraints, since it cannot entirely emulate the effects of hoover sealing.

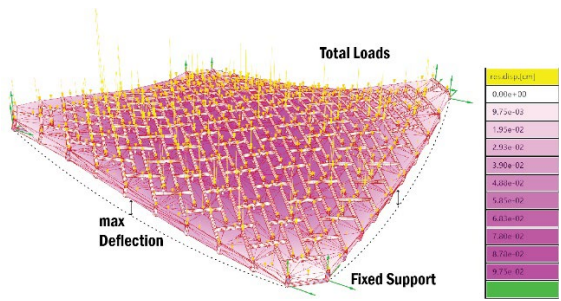

Figure 8: Load Distribution and Maximum Deflection Analysis of Rectangular Chainmail Shell Under Total Loads.

The structural evaluation is based on key international guidelines to ensure performance under various load conditions. SIA 262 outlines material properties and deflection limits (27), while SIA 261 defines categories such as dead, live, snow, and wind loads. These standards form the foundation for evaluating the adaptability and reliability of the chainmail shell structure (28).

Recycled PET (rPET) density varies with additives and processing. For simplicity, we use the maximum density of pure rPET for the shell structure (29):

$\rho = 1.13$ kN/m$^3$

The surface area (a) is: a=2m*2m=4m$^2$

Thickness(t): 0.08m

Deflection Analysis

The deflection limit is calculated using the formula:

Deflection Limit=L/250

where:

L=2m

Deflection Limit=2m/250=8mm

Load Analysis and Summary for Recyclable Plastic Shell Structure:

| Load Type | Formula | Constants | Results |

| Live Load (LL) | LL = max LL per unit area*a | max LL per unit area=0.4kN/m² | 1.60kN |
|---|---|---|---|
| Snow Load (SL) | SL = $\mu$*0.9kN/m²*a | $\mu$=0.8 (Maximum snow shape factor) | 2.88kN |
| Wind Load (WL) | WL = $C_p$*1.07kN/m²*a | $C_p$=0.75 (Maximum wind shape factor) | 3.21kN |
| Total Loads (TL) | TL = DL+LL+SL+WL | | |

## Form and Material Saving Optimization of Shell Structure

This phase (Figure 9) utilizes previous examinations, which illustrated the shell structure's ability to deform and carry loads, in the context of a specific design: a temporary, fast deployable shell structure. The maximum amplitude is established at 3 m to enable rapid assembly, providing practicality and easy access of installation.

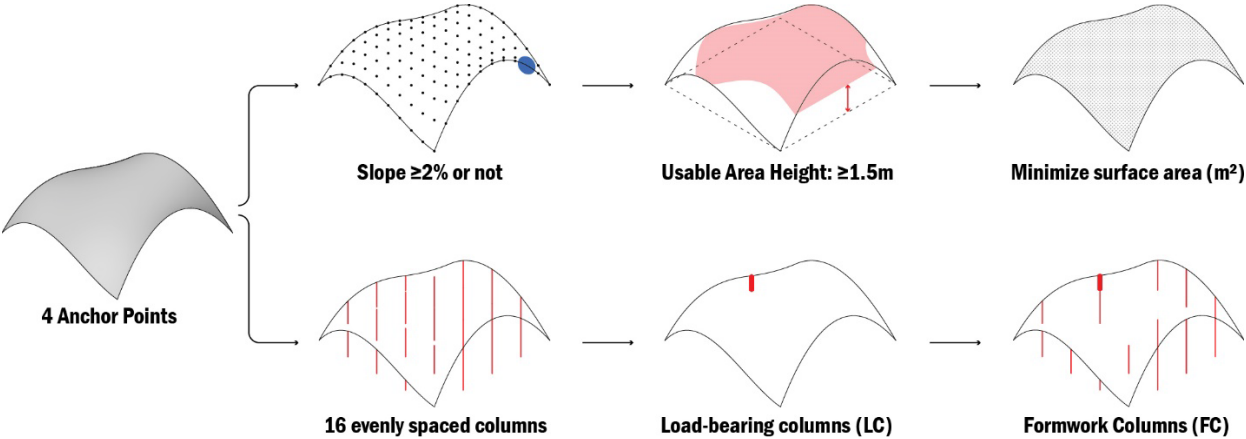

Figure 9: Design Workflow for Optimizing Anchor Point Configurations and Supporting Structures.

In order to optimize the shell's form and material configuration, we would like to focus on surface usability, structural support, and slope of roof. Materials for Load-Bearing Columns (LC) and Formwork Columns (FC), currently modeled as timber, are adaptable and can be substituted with sustainable, locally available alternatives to reduce costs and enhance flexibility.

Anchor points were tested in configurations of one, two (side or diagonal), three, and four to improve assembly and vacuum-sealing performance. Each configuration underwent 20 iterations, varying amplitude and control point positions, and was evaluated against four criteria. Maintaining an internal height of 1.5 meters, as per ISO 5912:2020 (30), ensured both usability and occupant comfort. Additionally, the roof design adhered to the requirements of IBC 2018 (31), which mandates a minimum drainage slope of 1/4 inch per foot (2% slope) for low-slope roofing systems.

To evaluate and ensure compliance with these slope requirements, the roof surface was subdivided into a 10×10 grid of points, as implemented in the Grasshopper workflow. For each grid point, normal vectors were derived, and the slope between neighboring points was calculated to verify that the overall roof geometry met or exceeded the 2% minimum slope. Iterations where any grid point's slope fell below this threshold were flagged with blue markers to visually indicate areas of non-compliance. Adjustments to control point positions and amplitudes were iteratively made, ensuring the roof curvature and drainage characteristics were optimized without compromising the overall form.

Using Karamba3D, the column system was initially designed with 16 evenly spaced columns, providing structural stability and sufficient interior space with a spacing of 0.5m. These columns were categorized into load bearing and formwork types, each assigned distinct support settings. Load-bearing columns were fully fixed at the base with fixed or pinned conditions at the top, while formwork columns were designed with sliding or pinned supports at the base and pinned supports at the top.

The optimization process focused on systematically reducing the number of formwork columns while ensuring that the resulting supporting surface maintained the same shape as the surface supported by the initial 16 columns. The filtering criteria for reduction were based on two key metrics: perimeter and surface area. Surfaces with negligible differences in these metrics were excluded, preserving forms that contributed most significantly to structural integrity. Through iterative adjustments, the number of formwork columns was minimized without compromising the ability of the combined load-bearing and remaining formwork columns to replicate the target shell geometry. This ensured that the optimized system maintained both efficiency and design consistency.

| Performance Metric | Formula | Objective |
| --- | --- | --- |
| Composite Chainmail Structure (CMS) | $CMS = \sum A_i$ | Minimize surface area (m²) |
| Usable Area (UA) | $UA = A_T - A_O$ | Maximize effective surface area (m²) |

| Load-bearing columns (LC) | $LC=\sum V_j$ | Minimize volume of fixed supports (m³) |
| --- | --- | --- |
| | | Minimize number of fixed supports |
| Formwork Columns (FC) | $FC=\min(\sum V_k)$ | Minimize volume of temporary supports (m³) |
| | | Minimize number of temporary supports |
| Slope (S) | $S = \Delta h/L$ | S≥2% to meet the standard |

This methodology offers a linear framework for determining an optimum free-form shell structure composed of rPET for temporary uses. Each phase enhances the design process, using rPET chainmail systems as a basis. The process defines experimental parameters, including amplitude, frequency, and part diameter, while using repeated testing and filtering to provide a clear advancement towards optimized outcomes.

The results follow these structured procedures, clarifying the consequences at each phase and demonstrating how the following steps refined the options. Each step of refinement is followed by analysis, explaining the logic for the selection of the final structure. This systematic approach facilitates the visualization of the rationale behind the ultimate decision, underlining its adaptability and efficiency for practical applications.

## Findings and Limitations

### Chainmail System Setup

With a consistent solid-to-gap ratio of 8%, the rectangular chainmail system demonstrated the best material efficiency, achieving the greatest deformation capacity with the least material usage. As shown in Figure 10, under identical solid-to-gap ratio conditions, the rectangular chainmail system exhibited the lowest overall weight, making it the most material-efficient choice for practical construction.

| Particle Designation | Particle Structure | Parameter | | 3*3 Particle Structure Show Joints | 3D print Sample Size | | | Solid-to-gap Ratio | Weight |
|---|---|---|---|---|---|---|---|---|---|
| | | L (mm) | d (mm) | | Length l(mm) | Width b(mm) | Height h(mm) | | |
| Triangular | 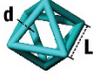 | 7.5 | 1 | 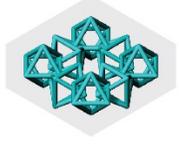 | 105 | 105 | 10 | 8% | 14g |
| Circular | 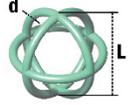 | 11 | 1 | 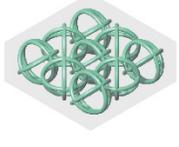 | 105 | 105 | 11 | 8% | 12g |
| Rectangular | 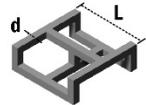 | 11 | 1 | 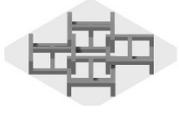 | 101 | 101 | 3.6 | 8% | 5g |

Figure 10: Geometric and Material Properties of Triangular, Circular, and Rectangular Chainmail Systems.

The thickness of individual rods varied across the three geometries—triangular, circular, and rectangular—significantly influencing deformation capacity. The thinner rods in the rectangular system allowed for greater deformation and adaptability, while the thicker rods in triangular systems increased stiffness but reduced flexibility. Notably, a limitation of this study is the lack of systematic exploration of how varying rod thicknesses within the same geometry affect deformation. Future work should address this by examining the relationship between rod thickness, deformation, and material efficiency within each geometry.

## 2D Sectional Testing

Figure 11 illustrates the enhanced deformation capacity of the rectangular chainmail system in comparison to triangular and circular systems. Exhibiting a maximum amplitude of 35 mm at a frequency of 9, it showcased enhanced flexibility and adaptability due to its linear form, which guarantees uniform load distribution and minimizes stress concentrations.

The rigid connections of the triangular system and the overlapping material of the circular system caused uneven stress distribution and diminished performance. The displacement of the rectangular system under vacuum-sealed circumstances highlights its efficiency in adaptive architectural applications, demonstrating its appropriateness for designs centered on deformation.

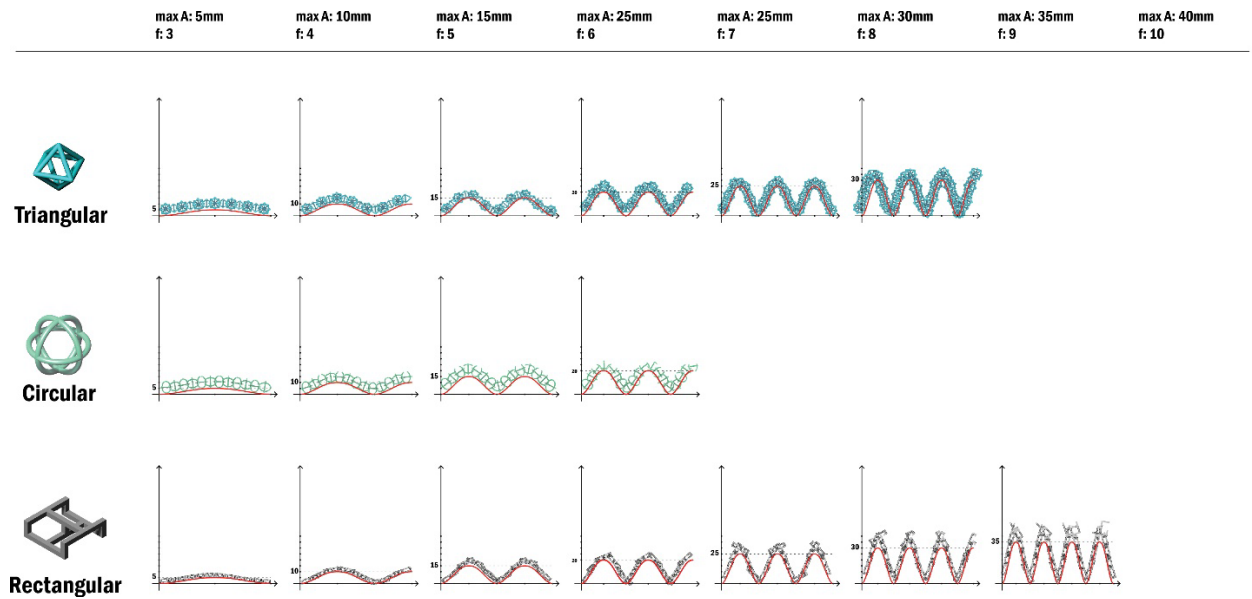

Figure 11: Deformation Responses of Triangular, Circular, and Rectangular Chainmail Systems Across Amplitude and Frequency Variations.

## 3D Shell Structure Generation: Exploring Rectangular Chainmail

Building on the results of the 2D testing, the rectangular chainmail system was selected for further analysis due to its superior performance. It not only exhibited the lightest weight but also achieved the highest deformation capacity, making it the most efficient and adaptable option for shell structure applications.

In this phase, a systematic exploration of Amplitude (A) and Frequency (f) was conducted, replicating the incremental testing approach from the 2D section testing. For each combination of A and f, 20 iterations of potential shell forms were generated. These iterations were represented in Plan view, with depth illustrated using black-and-white gradient images. As shown in the depth maps (Figure 12), darker colors indicate higher amplitude values, allowing for a clear visual assessment of deformation patterns across the generated iterations.

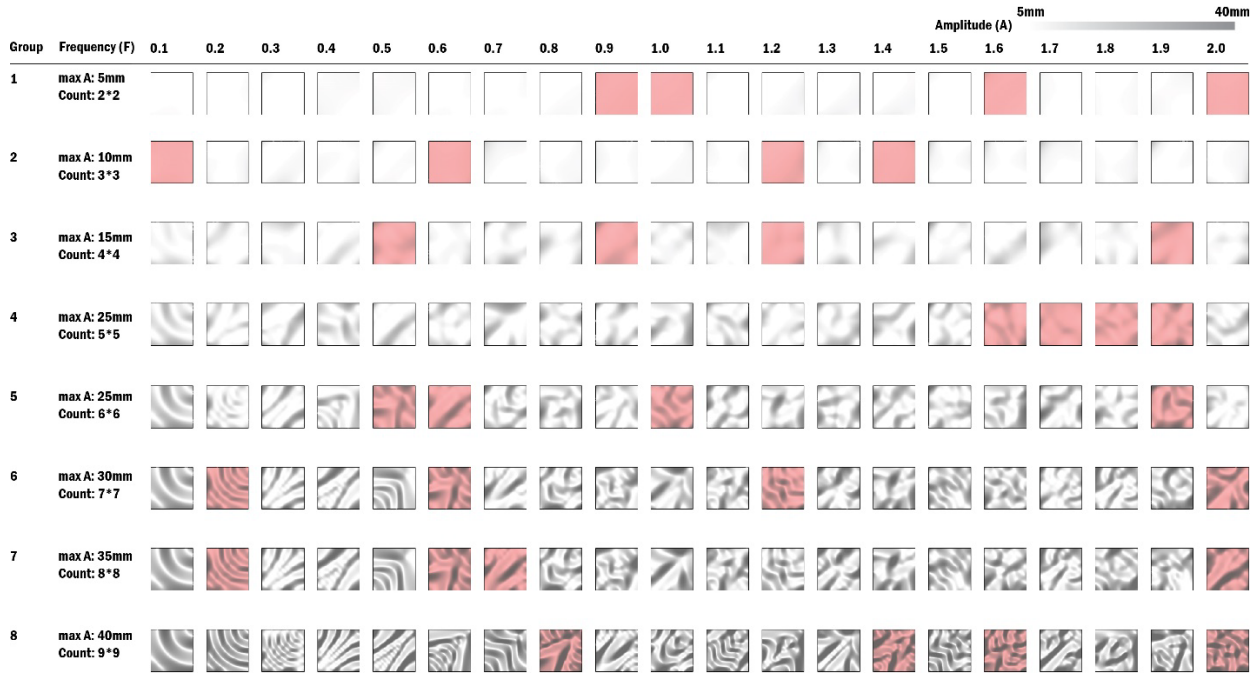

Figure 12: Amplitude and Frequency Variations in Rectangular Chainmail System.

## Physical Modeling and Vacuum Constraints

To validate the feasibility of the selected rectangular chainmail system under vacuum conditions, 1:20 scale physical models were created. Each model was tested against the formworks selected from Groups 1 to 5, simulating the effects of vacuum sealing. The objective was to assess whether the vacuum-sealed chainmail system could replicate the desired formwork in real-world conditions.

The results indicated that the system successfully fit the formworks for Groups 1 through 4, with Group 4 (Max A:25mm, F:6) achieving the highest deformation under vacuum constraints. However, for Group 5, certain areas failed to conform to the intended geometry, as highlighted by red-circled discrepancies in a series of freeze-frame images (Figure 13). These mismatched regions indicate that the deformation capacity of the system under vacuum is limited to the parameters of Group 4.

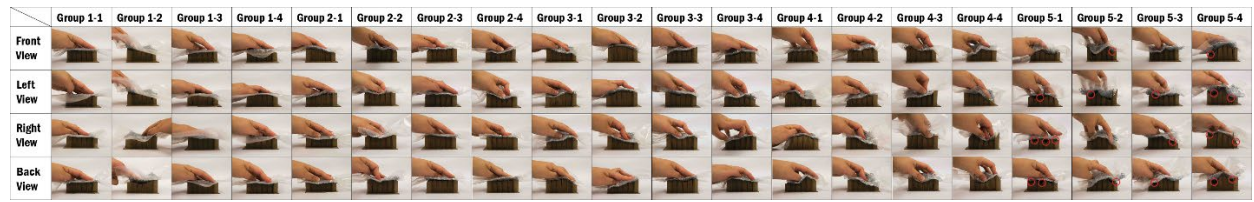

Figure 13: Vacuum-Sealed Rectangular Chainmail System: Physical Model Testing Across Groups and Views.

## Load Analysis for Free-Form Shell Structure

This phase evaluates the structural feasibility of the 16 iterations selected from Group 1 to Group 4, using Karamba3D for simulation. The analysis considered four load types—dead, live, snow, and wind. These simulations aimed to assess whether the deformation of the iterations remains within the permissible deflection limit of 8 mm, as established in the methodology.

Figure 14 confirms that all tested iterations meet the deflection criteria, with surfaces exhibiting greater deformation achieving reduced displacements. This highlights the relationship between geometric flexibility and structural stability, emphasizing the importance of optimized surface shapes in minimizing displacement and enhancing performance.

| Model | Dead Load (kN) | Live Load (kN) | Snow Load (kN) | Snow Load (kN) | Net Loads (kN) | Net Loads (kN) | Max Displacement (mm) |
|---|---|---|---|---|---|---|---|
| Group 1-1 | 0.10 | 1.60 | 2.88 | 2.88 | 3.21 | 4.58 | 5.30 |
| Group 1-2 | 0.10 | 1.60 | 2.88 | 2.88 | 3.21 | 4.58 | 5.64 |
| Group 1-3 | 0.10 | 1.60 | 2.88 | 2.88 | 3.21 | 4.58 | 5.77 |
| Group 1-4 | 0.10 | 1.60 | 2.88 | 2.88 | 3.21 | 4.58 | 5.11 |
| Group 2-1 | 0.12 | 1.60 | 2.88 | 2.88 | 3.21 | 4.60 | 4.64 |
| Group 2-2 | 0.12 | 1.60 | 2.88 | 2.88 | 3.21 | 4.60 | 4.83 |
| Group 2-3 | 0.12 | 1.60 | 2.88 | 2.88 | 3.21 | 4.60 | 4.61 |
| Group 2-4 | 0.12 | 1.60 | 2.88 | 2.88 | 3.21 | 4.60 | 4.02 |
| Group 3-1 | 0.14 | 1.60 | 2.88 | 2.88 | 3.21 | 4.62 | 3.22 |
| Group 3-2 | 0.14 | 1.60 | 2.88 | 2.88 | 3.21 | 4.62 | 3.32 |
| Group 3-3 | 0.14 | 1.60 | 2.88 | 2.88 | 3.21 | 4.62 | 3.65 |
| Group 3-4 | 0.14 | 1.60 | 2.88 | 2.88 | 3.21 | 4.62 | 3.53 |
| Group 4-1 | 0.16 | 1.60 | 2.88 | 2.88 | 3.21 | 4.64 | 2.92 |
| Group 4-2 | 0.16 | 1.60 | 2.88 | 2.88 | 3.21 | 4.64 | 0.93 |
| Group 4-3 | 0.16 | 1.60 | 2.88 | 2.88 | 3.21 | 4.64 | 1.65 |
| Group 4-4 | 0.16 | 1.60 | 2.88 | 2.88 | 3.21 | 4.64 | 1.21 |

Figure 14: Load and Displacement Analysis of Grouped Rectangular Chainmail Models Under Various Load Conditions.

This demonstrates that the rectangular chainmail system, when vacuum-sealed and subjected to regulated loading conditions, maintains its structural integrity and is feasible for practical applications. The analysis supports the system's ability to accommodate different levels of deformation while adhering to safety standards.

## Form and Material Optimization of Temporary Shell Structure

The optimization process categorized the iterations into five groups based on the number and placement of anchor points (AC), as shown in Figure 15. Each category contained 20 iterations, evaluated against four key criteria: CMS (Composite Chainmail System area), UA (Usable Area), LC (Load-bearing Columns), and FC (Formwork Columns). Rankings for these criteria were assigned on a 1–100 grade, where 100 represents the best performance and 1 the worst, ensuring a standardized evaluation across all iterations.

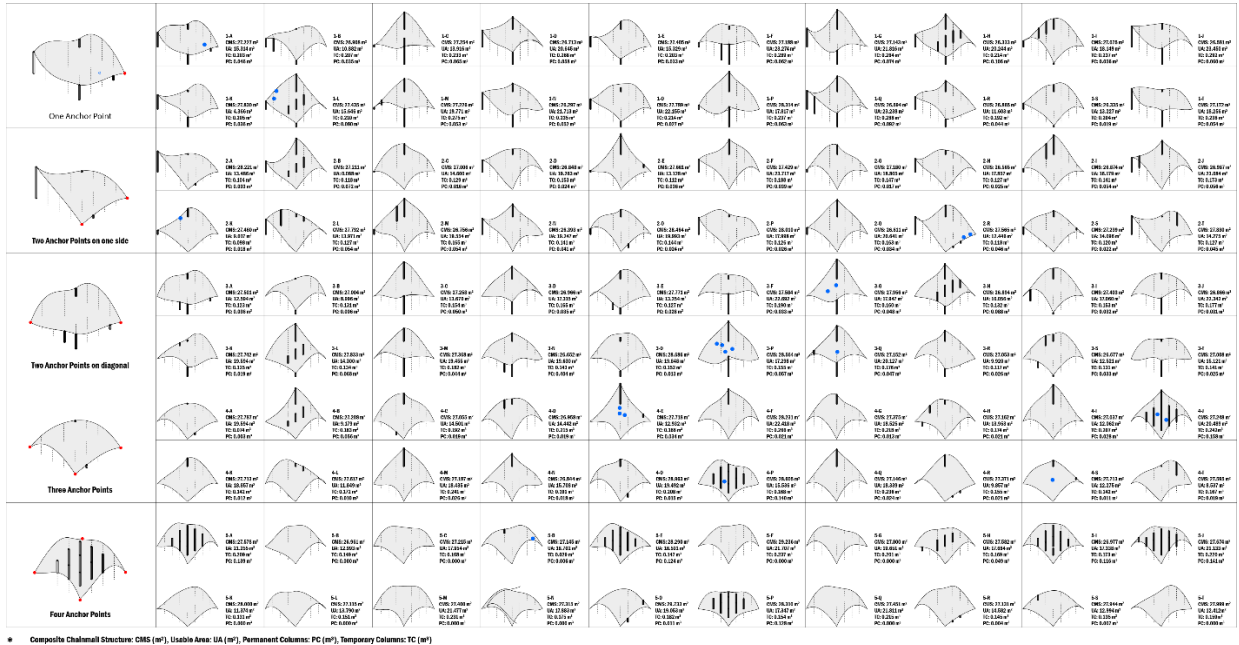

Figure 15: Anchor Point Configurations and Performance Analysis for Composite Chainmail Structures.

Figure 16 clearly illustrates the performance rankings of various anchor point configurations, highlighting the weighted contributions of essential criteria. CMS and UA, each weighed at 0.4, were selected to emphasize material efficiency and the maximizing of functional space, underscoring their significance in design optimization. LC and FC, each weighed at 0.1, were assigned secondary importance due to their diminished relevance in temporary structural applications.

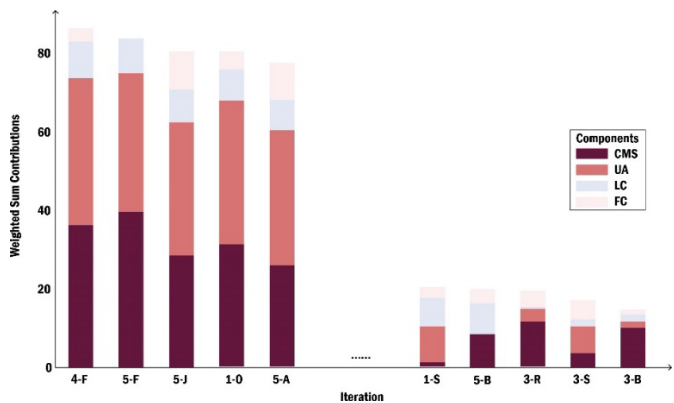

Figure 16: Weighted Performance Contributions of Anchor Point Configurations Across Key Criteria.

Before using this weighted scoring system, iterations that did not satisfy the drainage criterion of a least 2% roof slope, as stipulated by (31), were omitted to ensure adherence to water drainage regulations. The subsequent iterations were ranked according to their weighted ratings, which integrated the four criteria: CMS, UA, LC, and FC. This method determined the most effective configurations, with the highest-ranked iterations attaining an optimal equilibrium among material efficiency, functionality, and structural viability.

The findings (Figure 17) emphasize setups with reduced CMS for material conservation and elevated UA to optimize usable area, while ensuring low LC and FC values to reduce structural and assembly requirements. The iterative method guaranteed the retention of just the most efficient and useful designs for subsequent study, enabling a thorough assessment of their practical application potential. The net loads were applied to this structure, resulting in a maximum displacement of 3.56 mm, indicating its structural integrity and practical viability.

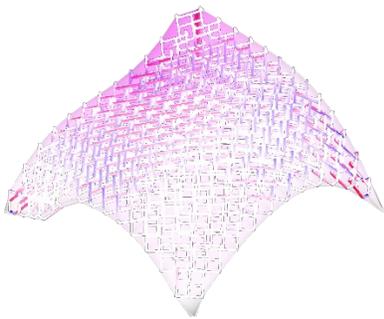

Figure 17: Optimized Structural Configuration with Minimal Material Usage and Maximum Usable Area.

## Discussion/Future Work

Building on the findings from the previous tests and studies, the composed chainmail structure demonstrates unique advantages, combining adaptability, load-bearing capacity, and ease of assembly. These attributes make it highly suitable for deployment in extreme conditions and remote locations, where accessibility and transportation costs are significant challenges. For scenarios requiring specialized transport methods, such as helicopters, submarines, or spacecraft, the lightweight and modular nature of the chainmail structure is especially advantageous.

While the structure has been validated as an effective temporary shell structure for providing post-disaster shelter, its potential extends beyond this application. With further development, the composed chainmail structure could be adapted for diverse, high-performance applications in extreme environments. However, these applications will require additional considerations, such as the structure's performance under cosmic radiation, extreme temperature fluctuations, and other environmental stresses. For instance, while the use of recyclable plastic materials is sustainable and efficient, these materials may not meet the demands of extreme conditions without further enhancement.

In such cases, additive processes could be employed to incorporate specialized substances into the plastic, enhancing its properties to withstand harsher environments. Below, we outline several future applications and the criteria that must be addressed for successful implementation:

1. Space Debris Shielding
   The composed chainmail structure could serve as a lightweight, adaptable shield against space debris. However, its deployment in outer space requires addressing radiation resistance and thermal durability to endure cosmic radiation and extreme temperature variations. Additive processes could integrate radiation-blocking materials or thermal stabilizers to improve their performance in this context.

2. High-Wind Rescue Structures
   In high-altitude or cliffside rescue scenarios, chainmail systems could be used to create temporary shelters or structural supports. However, wind resistance is critical in these settings. Future work should integrate CFD simulations to optimize aerodynamic stability and consider materials with enhanced rigidity to counteract strong winds.

3. Underwater Platforms
   For high-pressure underwater environments, chainmail structures could serve as exploration or rescue platforms. Current recyclable plastics may deform or lose integrity under prolonged high-pressure conditions. Incorporating pressure-resistant additives into the plastic could significantly improve performance, enabling use in underwater exploration and rescue missions.

# Conclusion

## Summary of Contributions

This paper introduces two significant contributions to advancing sustainable construction by leveraging recycled plastic filaments and modular chainmail systems, bridging a critical gap in material circularity and architectural adaptability.

1. A design methodology for integrating recycled plastic filaments into chainmail systems optimized for vacuum-sealed applications. This methodology encompasses computational workflows for 2D sectional testing, 3D shell structure generation, and physical modeling, alongside structural optimization for deformation capacity and material efficiency. It demonstrates the adaptability of chainmail geometries under vacuum constraints and their suitability for architectural applications, from temporary shelters to lightweight structural systems.

2. An optimization strategy for form and material configuration in temporary shell structures, focusing on practical deployment. This strategy includes a systematic evaluation of performance metrics such as material usage, load distribution, and water drainage efficiency, as demonstrated by the best-performing rectangular chainmail configuration. The findings highlight the feasibility of integrating recyclable plastic materials into scalable, deployable designs that meet international standards.

## Potential Impact

This study presents a material-driven biomimetic strategy that transitions from traditional form-based biomimicry to an approach focused on adjustable material properties for architectural applications. This research creates a systematic framework that connects experimental material science with architectural design, demonstrating the viability of incorporating recycled polyethylene terephthalate (rPET) with vacuum-sealed chainmail fabrics for deployable shelter structures. The results offer an innovative technical approach and theoretical foundation for sustainable material use in architecture, illustrating the effective upcycling of adaptive materials into high-performance building elements.

While the primary application focuses on temporary shelter solutions, the adaptability of the proposed methodology suggests potential extensions to other architectural systems, such as disaster-resilient structures, lightweight façades, and modular building components. The structural adaptability and recyclability of vacuum-sealed chainmail fabrics expand potential applications in extreme environments, such as underwater habitats and extraterrestrial construction. This study establishes a foundation for scalable applications of adaptive materials in sustainable construction by integrating computational design, material circularity, and performance-oriented material selection, providing a replicable model for future architectural advancements.